\documentclass{osa-article}

\usepackage{amsbsy}
\usepackage{amsfonts}
\usepackage{commath}
\usepackage{graphicx}
\usepackage{epstopdf,mathrsfs}
\usepackage{booktabs}
\usepackage{bbold}
\usepackage{epsfig}
\usepackage{dcolumn}
\usepackage{bm}
\usepackage{braket} 
\usepackage{CJK}
\usepackage{multicol}
\usepackage{color,soul}
\usepackage{float}
\usepackage{lineno}
\usepackage{chngcntr}

\journal{oe}

\newcommand{\new}[1]{#1}

\articletype{Research Article}

\begin{document}

\title{Single-shot characterization of vector beams by generalized measurements}

\author{M. A. Al Khafaji,\authormark{1,2,*} C. M. Cisowski,\authormark{1} H. Jimbrown,\authormark{1} S. Croke,\authormark{1} S. P\'{a}dua,\authormark{1,3} and S. Franke-Arnold,\authormark{1}}

\address{\authormark{1}School of Physics and Astronomy, University of Glasgow, G12 8QQ, Glasgow, Scotland, UK\\
\authormark{2}Fraunhofer CAP, G1 1RD, Glasgow, Scotland, UK\\
\authormark{3}Departamento de F\'{\i}sica, Universidade Federal de Minas Gerais, 31270-901 Belo Horizonte, Minas Gerais, Brazil}

\email{\authormark{*}m.al-khafaji.1@research.gla.ac.uk} 

\begin{abstract}
Vector vortex beams, featuring independent spatial modes in orthogonal polarization components, offer an increase in information density for emerging applications in both classical and quantum communication technology. Recent advances in optical instrumentation have led to the ability of generating and manipulating such beams. Their tomography is generally accomplished by projection measurements to identify polarization as well as spatial modes. In this paper we demonstrate spatially resolved generalized measurements of arbitrary vector vortex beams. We perform positive operator valued measurements (POVMs) in an interferometric setup that characterizes the vector light mode in a single-shot. This offers superior data acquisition speed compared to conventional Stokes tomography techniques, with potential benefits for communication protocols as well as dynamic polarization microscopy of materials.  
\end{abstract}

\section{Introduction}

Polarization is one of the key properties of light, affecting the propagation and scattering of optical beams and providing a convenient degree of freedom for classical and quantum communication. While laser beams typically have homogeneous polarization profiles, recent years have seen an increasing interest in light designed with spatially varying polarization structures, including vector vortex beams \cite{cylvortex,spontvortex,qplatevortex,cylmathvect,singvortex09}, Poincar\'e beams \cite{Beckley:10,Beckley:12,delin19} and even skyrmionic beams \cite{Gao2020,lin21}.  These vector beams are characterized by \new{non-separable} correlations between their polarization and spatial degree of freedom \cite{fickler14,balthazar16}, with intriguing applications for quantum \new{\cite{DAmbrosio2012,Parigi2015,ambrosio16,sit2017qkd}} and quantum inspired protocols \cite{milione15,li16,Berg-Johansen15,zhao15,jia:19, Kopf21,toppel14}.

Moreover, polarization plays an essential role for various metrological applications, including pharmaceutical and biological microscopy \cite{sparks09,angelo19}, ellipsometry \cite{garcia-caurel13,azzam19,Azzam1977}, and stress analysis of materials \cite{hearn97,daniels17}. In these cases, the polarization profile of a probe beam is modified during transmission through a medium, and analysis of the resulting polarization structure allows \new{one} to test its optical activity \cite{Azzam1977,schubert05,Losurdo2013,Hinrichs2014,Ohlidal:20}. An accurate determination of the spatially dependent polarization is therefore an important task in classical as well as quantum optics \cite{toussaint04,Rudnicki20}. 

The polarization degree of freedom can be represented in a two-dimensional Hilbert space, making it an ideal realisation for a qubit.  A state belonging to a Hilbert space of dimension $d$ demands $d^2$ measurements for the state reconstruction \new{\cite{caves2002unknown}}. Tomography of the local polarization therefore requires at least four measurements, which can be associated with the four real parameters of the Stokes vector \cite{james01,altepeter05,focus13,Foreman2015,Toninelli19}.  For convenience, these parameters are usually determined from overcomplete measurements,  performed as projections on the mutually unbiased basis states, detecting the light intensity of the horizontal, vertical, diagonal, anti-diagonal, right and left hand polarization components.  

Generalized measurements \cite{chuang10,barnett09,clarke01} provide a convenient strategy for realizing full tomography of quantum states with a minimum number of measurements \cite{singapore04,Ling2006,Pimenta10,pimenta13,bent15, Dada2019}. 
The operators that define these measurements can be chosen to form a minimum informationally complete positive operator value measure (MIC-POVM) \cite{renes04,paiva10}.  In the case of polarization tomography, this entails a reduction from 6 to 4 measurements, with obvious benefits in reducing the required time when performing sequential measurements, or in increasing efficiency when performing parallel measurements, which becomes crucial for applications with low intensity light or single photons. 
Experiments using MIC-POVMs for quantum state tomography have so far been realised only for homogeneously polarized light \cite{ling2006optimal}. 

In this paper we demonstrate {\it spatially resolved} MIC-POVMs for vector beams. An interferometric setup allows us to couple the polarization degree of freedom to the path (or linear momentum) degree of freedom via a polarizing beamsplitter, perform unitary operations in the combined state space of polarization and path degree of freedom, associate the four POVM states with four outputs of our interferometer, and finally to perform spatially resolved generalized measurements with these POVM states. 

Our technique can fully characterize the inhomogeneous transverse polarization structure of arbitrary vector beams.
We test our setup on several vector beams that we design using a DMD-based vector mode generator \cite{Mirhosseini13,Mitchell16,selyem19,rosales2020}, and compare its performance against tomography based on spatially resolved Stokes measurements \cite{selyem19,DErrico2017,Ndagano16,Ureta18}. 
We conclude by outlining potential applications for one-shot polarimetry of photo-active materials.  


\section{Theory}

The polarization quantum state of a photon, or the polarization of a classical light mode, is fully characterized by four parameters. A convenient choice are the four real parameters of the Stokes vector,
\begin{equation} \label{Stokes}
   {\bf S} = \bigl(S_{0},  S_{1},  S_{2},   S_{3} \bigr)^{T} = \bigl(\mathrm{I_H} +\mathrm{I_V},  \mathrm{I_H} -\mathrm{I_V},  \mathrm{I_D}- \mathrm{I_A},   \mathrm{I_R}-\mathrm{I_L} \bigr)^{T}, 
\end{equation}
which can be readily identified from a set of six intensity measurements, where ${\rm I}_j$, for $j \in \{ {\rm H, V, D, A, R, L}\}$ 
represents the intensity in the horizontal, vertical, diagonal, antidiagonal, right and left polarization component respectively. Normalizing with the total intensity $S_0$ yields,  
\begin{equation}
    {\bf S}^{\rm(N)}=\left(S_0, S_1, S_2, S_3  \right)^T /S_0.
    \label{NStokes}
\end{equation}
The latter three components form the reduced Stokes vector which is commonly used to describe the polarization state in the Poincar\'e sphere. 
Homogeneously polarized light is described by a single vector ${\bf S}$. Vectorial light beams, such as vector vortex beams, or Poincar\'{e} beams, instead, have a non-homogeneous polarization structure, so that ${\bf S}({\bf r}_{\bot})$ varies as a function of transverse position ${\bf r}_{\bot}=(x,y)$. 

Adopting a quantum language, we may write a paraxial light beam (or its photon wavefunction) at a given time as,
\begin{equation}
    \ket{\psi}= \ket{u_{\rm H}}\ket{H} +e^{i\phi}\ket{u_{\rm V}}\ket{V},
    \label{state}
\end{equation}
where $\ket{H}$ and $\ket{V}$ are the orthonormal linear polarization states in the horizontal and vertical directions, respectively; $\ket{u_j}$  for $j \in \{ {\rm H} , {\rm V} \} $ are transverse spatial states (which need not be orthogonal, or normalized), and $\phi$ is a constant phase. The local polarization state at transverse spatial position ${\bf r}_{\bot}$ is thus given by,

\begin{equation}
    \ket{\psi_p({\bf r}_{\bot})} = \braket{{\bf r}_{\bot}|\psi} = u_{\rm H}({\bf r}_{\bot}) \ket{H} + e^{i \phi} u_{\rm V}({\bf r}_{\bot}) \ket{V},
\end{equation}
where $u_j({\bf r}_{\bot})= \braket{{\bf r}_{\bot}|u_j}$ are the complex amplitudes of the transverse light profile described by $\ket{u_j}$. The aim of spatially resolved polarimetry is to characterize these local polarization states, across the transverse beam profile. This has previously been achieved through Stokes measurements, using overcomplete intensity measurements in the six polarization states with a camera after suitable polarization filters \cite{selyem19}. A single shot method for obtaining the polarization state of a light beam is the four-detector scheme \cite{Azzam1977,Ling2006} which allows the measurement of the state polarization with four detectors at the exit of an interferometer. 


We use a quantum language to emphasise the strong connection with quantum tomography, interpreting the polarization measurement as a POVM. A POVM is a set of operators $\{ \hat{\pi}_i \}$ satisfying $\hat{\pi}_i \geq 0$, and the completeness condition $ \sum_{i} \hat{\pi}_i=\hat{I}$, where $\hat{I}$ is the identity operator \cite{chuang10,barnett09}. The probability of obtaining outcome $i$ in a measurement on a quantum system prepared in state $\ket{\psi}$ is given by,
\begin{equation}
 \label{P}
{\rm P}_i = \bra{\psi} \hat{\pi}_i \ket{\psi}.     
\end{equation}

\begin{figure}[H]
    \centering
    \includegraphics[width=0.9\linewidth]{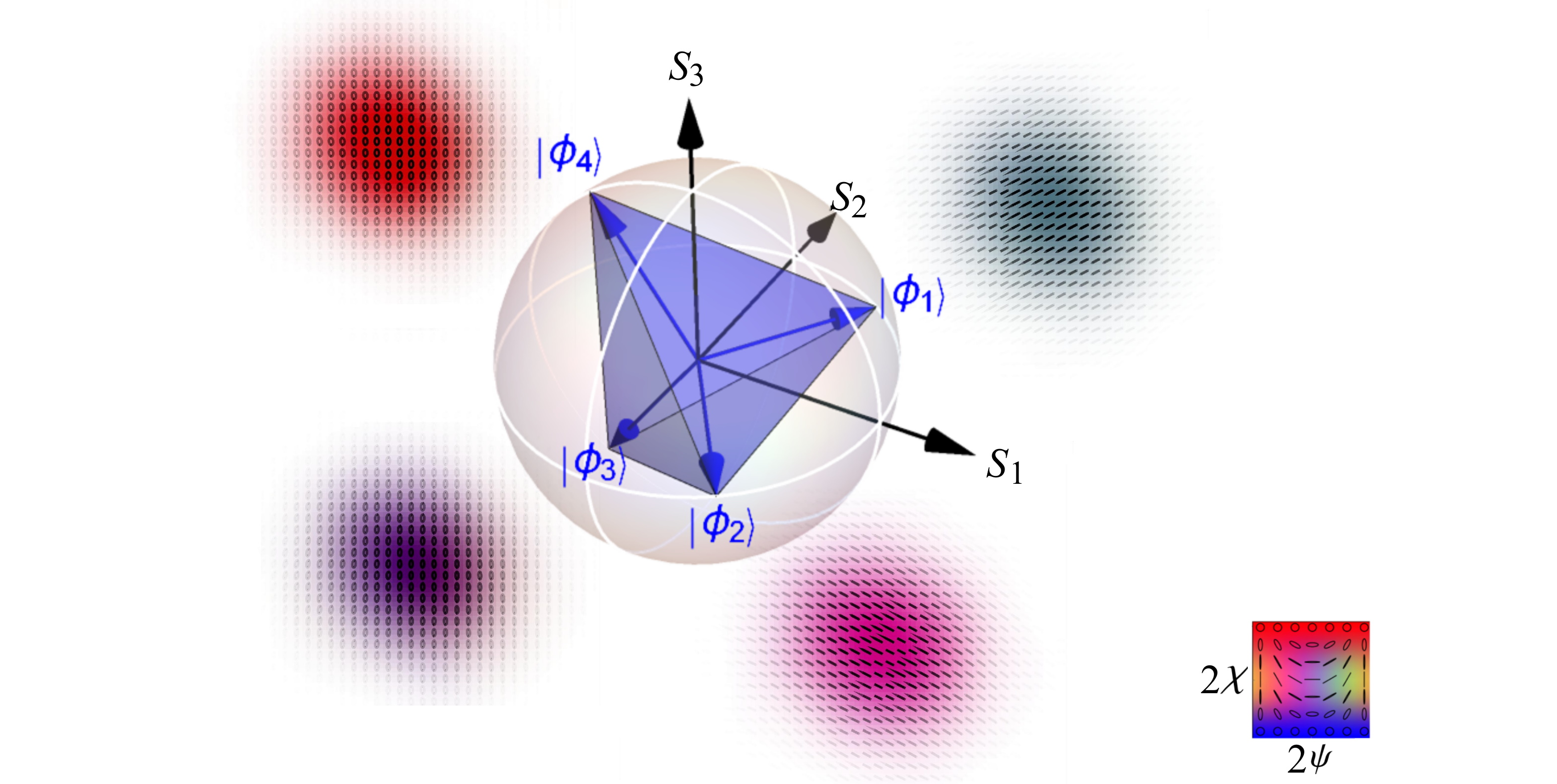}
    \caption{The POVM states $\ket{\phi_1},\ket{\phi_2},\ket{\phi_3},\ket{\phi_4}$ form a tetrahedron on the Poincar\'e sphere. 
    Their polarization distributions are extracted from experimental Stokes measurements. The polarization ellipses are plotted in black, the coloured background, modulated in opacity according to the beam intensity, provides a continuous mapping of the ellipticity $\chi$ and orientation $\psi$ of the polarization ellipse at each pixel, as illustrated in the inset.}
    \label{fig:poincare}
\end{figure}

A particularly appealing form of generalized measurement is the symmetric informationally complete POVM (SIC-POVM) \cite{renes04}, where the POVM elements are positive multiples of projectors onto pure states with equal pairwise overlap between different POVM elements. The $d^{2}$ elements of the set then have the form $\hat{\pi}_i = \frac{1}{d}\ket{\phi_i}\bra{\phi_i}$ with $|\bra{\phi_i}\phi_j\rangle|^2 = (d\delta_{ij} +1)/(d+1)$, where $\delta_{ij}$ is the Kronecker delta. For $d=2$, the states $\{ \ket{\phi_i} \}$ form the vertices of a tetrahedron in the Bloch (or Poincar\'e) sphere as illustrated in Fig.~\ref{fig:poincare}. As the states are equally spaced and symmetrically distributed on the Poincar{\'e} sphere they do not privilege the reconstruction of any particular qubit state \new{and give the same fidelity of reconstruction for any input state \cite{clarke01}}. In our work we use the set of POVM states introduced by \cite{Ling2006}, 

\begin{eqnarray}\label{eqpovm}
        & \ket{\phi_1}= a\ket{H} + b\ket{V}, \qquad  
        & \ket{\phi_2}= a\ket{H} - b\ket{V}, \nonumber \\
        & \ket{\phi_3}= b\ket{H} + ia\ket{V}, \qquad 
        & \ket{\phi_4}= b\ket{H} - ia\ket{V}, 
\end{eqnarray}
where $a = \sqrt{\frac{1}{2}+\frac{1}{2\sqrt{3}}}$ and $b = \sqrt{\frac{1}{2}-\frac{1}{2\sqrt{3}}}$. 

\begin{figure}[H]
    \centering
    \includegraphics[width=0.9\linewidth]{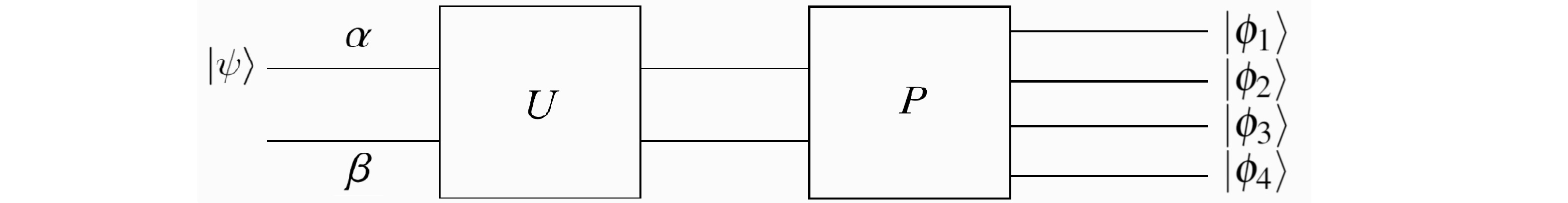}
    \caption{Schematic representation of our POVM measurement scheme. The input polarization state $\ket{\psi}$ enters the apparatus in path $\alpha$. An additional path labelled $\beta$ realizes an extension of the state space to allow a four-outcome measurement. The unitary $U$ acts on both polarization and path degrees of freedom, and the final projective measurement in the extended hilbert space is represented by $P$. $\ket{\Phi_i}$ are the POVM states that define the POVM elements.}
    \label{fig:theoryfig}
\end{figure}

A common strategy to perform 
measurements on the non-orthogonal POVM states $\ket{\phi_i}$ is to increase the Hilbert space of the input states by adding an auxiliary state, forming a so-called Naimark extension of the measurement \cite{chuang10,barnett09}. In common with previous works \cite{clarke01,Ling2006,ling2006optimal}, we choose the path degree of freedom. We suppose the photon in the state Eq.~(\ref{state}), \new{illustrated in Fig.\ref{fig:theoryfig}} has two propagation directions available, characterized by the photon's linear momentum $\ket{k_\nu}$ along the path $\nu$ = $\alpha, \beta$, so that the photon state is now written in terms of the extended basis $\{\ket{H}\otimes \ket{k_\nu}, \ket{V}\otimes \ket{k_\nu}\}$. By applying an appropriately chosen unitary $U$ to the joint state, followed by a projective measurement \new{$P$}, we can implement the desired POVM. Full details, which follow \cite{Ling2006} are given in the Supplementary Documentation. 

In experiments with classical light, the detectors register intensities, or for our spatially resolved light beams, intensity profiles. The probabilities derived in the single photon POVM picture then become the normalized intensity profiles at each detector. In our experiment we record the normalized four component vector, 

\begin{equation}
\label{I}
    {\bf I}^{\rm (N)} = (I_1,\ I_2, \ I_3, \ I_4)^T /I_{\rm t},
\end{equation}
where $I_i$ are the measured intensities at the interferometer outputs and $I_t= \sum^{4}_{i = 1} I_i$. 

The four measured intensities are sufficient to completely determine the state, or its associated Stokes vector. This can be seen by rewriting the expectation value of Eq.~(\ref{P}) as a trace over the density operator, so that  
$P_i=I^{\rm (N)}_i= {\rm Tr}( \hat{\pi}_i \hat{\rho})$. Expressing the density operator in terms of the normalized Stokes vector from Eq.~(\ref{NStokes}), we can write \cite{altepeter05},
\begin{equation}
\hat{\rho} = \frac{1}{2}
\begin{pmatrix}
        1 + S^{\rm (N)}_1 &   S^{\rm (N)}_2-i S^{\rm (N)}_3 \\
         S^{\rm (N)}_2+i S^{\rm (N)}_3 & 1-S^{\rm (N)}_1
    \end{pmatrix}
\end{equation}
 and obtain the matrix equation,
    \begin{equation}
    \label{instrument_eqn}
    {\bf I}^{\rm (N)} = \Pi \cdot {\bf S}^{\rm (N)},
\end{equation}
where the $4 \times 4$ matrix $\Pi$ is known as the instrumentation matrix \cite{Azzam88, Ling2006}. By inverting this equation, we can obtain the normalised Stokes vector at each position $\bf{r}_\bot$ from the intensity data obtained for the 4 POVM elements, with spatially dependent intensity measurements yielding spatially dependent Stokes measurements. 



For an ideal polarimeter that experiences no losses, a possible instrumentation matrix is,
\begin{equation}
    \label{ideal_instrument}
    \Pi = \frac{1}{4}\begin{pmatrix}
        1 & \sqrt\frac{1}{3} & \sqrt\frac{2}{3} & 0 \\
        1 & \sqrt\frac{1}{3} & -\sqrt\frac{2}{3} & 0 \\
        1 & -\sqrt\frac{1}{3} & 0 & -\sqrt\frac{2}{3} \\
        1 & -\sqrt\frac{1}{3} & 0 & \sqrt\frac{2}{3}
    \end{pmatrix},
\end{equation}  
where the horizontal entries of the instrumentation matrix are given by the Stokes vectors of the individual POVM states $\ket{\phi_i}$.  We show in the next section how this instrument matrix is realised by 
the configuration of optical equipment in our experimental setup. 

\section{Experimental realisation}

\begin{figure}[H]
    \centering
    \includegraphics[scale=0.85]{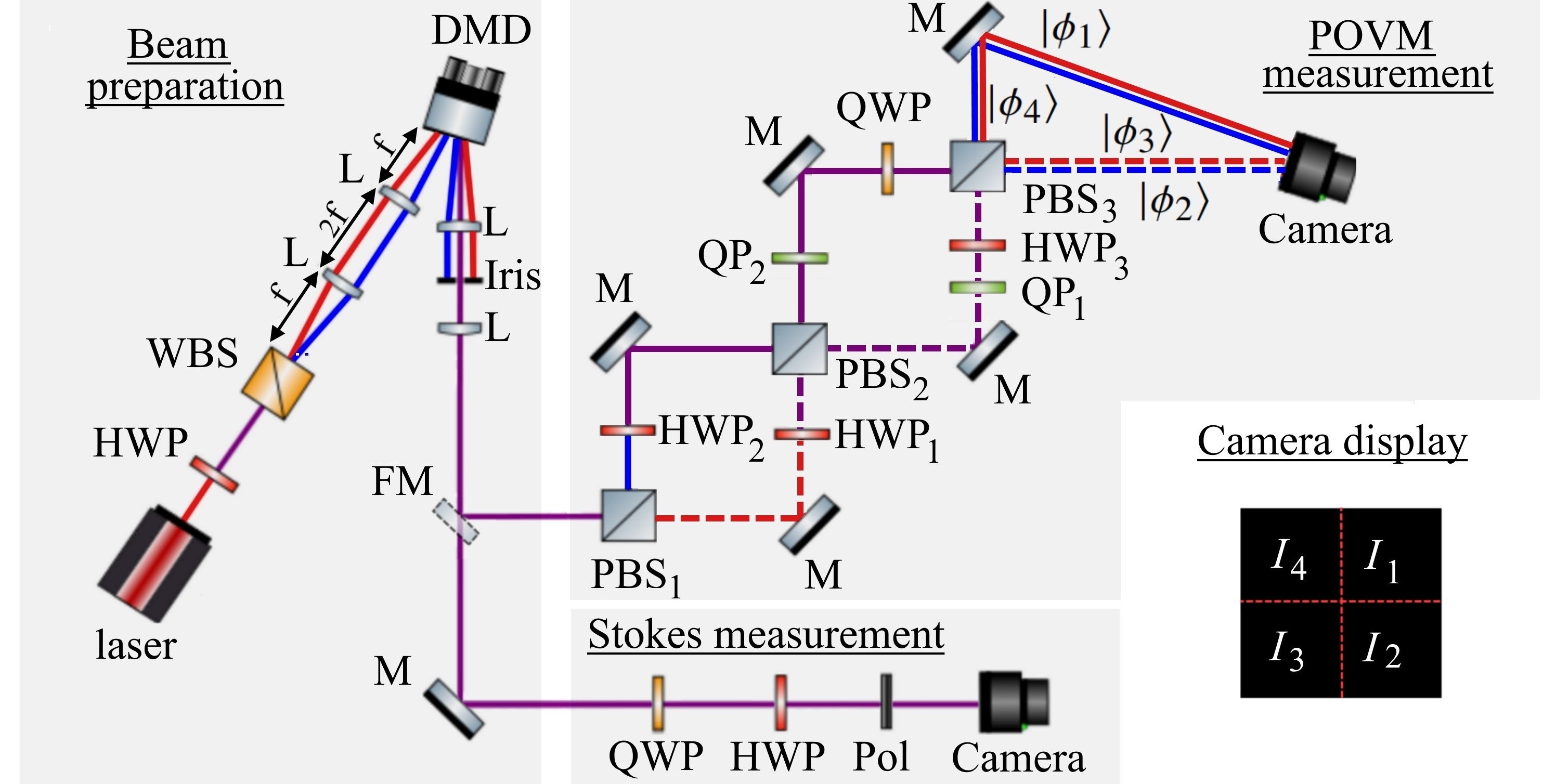}
    \caption{Experimental setup for 
    spatially dependent POVM measurements. The colour of the beams indicates the polarization state, red for H, blue for V and purple for a superposition of both. Between ${\rm PBS}_1$ and ${\rm PBS}_3$, dotted lines represent the $\alpha$ path whereas solid lines represent the  $\beta$ path. The optical setup shows half waveplates (HWP), quarter waveplates (QWP), quartz plates (QP), polarizers (Pol), Wollaston beam splitter (WBS), lenses (L) with focal length f, mirrors (M) and a flip mirror (FM). The inset shows the four quadrants of the camera display, recording the projection measurements onto the four POVM states. \new{Note that the incidence angle of the states $|\phi_1\rangle$ and $|\phi_4\rangle$ on the camera is around 0.1 radians, but is exaggerated here for clarity.}}
    \label{fig:exp_setup}
\end{figure}

We demonstrate and evaluate spatially dependent POVM measurements using the setup shown in Fig.~\ref{fig:exp_setup}. Our POVM measurement follows the general outline of a polarimeter for homogeneous polarisation states in \cite{Ling2006}, however applied to beams with spatially varying polarization profiles.

We generate arbitrary vector beams of the form Eq.~(\ref{state}), using the technique introduced in \cite{selyem19}, based on a digital micromirror device (DMD). A half-waveplate followed by a Wollaston beam splitter (WBS) separates a linearly polarized HeNe-laser beam into equal vertical and horizontal polarization components. A telescope system directs both polarization components onto a DMD, where a multiplexed hologram  independently shapes their amplitude and phase. 
The two spatially shaped polarization components are superimposed in their respective first diffraction order after diffraction off the DMD hologram and the desired order is spatially selected using an iris. A pair of lenses ensures that the generated beam is magnified and collimated upon further propagation. 

A flip mirror (FM) directs the beam either to our spatially resolved POVM measurement setup or to a conventional Stokes polarimetry setup, indicated by the different panels in Fig.~\ref{fig:exp_setup}. We implement full Stokes tomography as described in \cite{Toninelli19} by performing a sequence of six intensity measurements $I_j({\bf r}_\bot)$, for $j \in \{ {\rm H, V, D, A, R, L}\}$, using a rotating quarter-waveplate (QWP), half-waveplate (HWP), polarizer (Pol) and recording the intensity on a CMOS camera (Thorlabs DCC1645C). This six-measurement Stokes tomography provides a control against which we assess the performance of our POVM tomography. We use gold-plated rather than dielectric mirrors throughout the setup to reduce unwanted polarization transformations. 

The state entering our POVM measurement scheme is, according to Eq.~(\ref{state}), together with the labelling of input path $k_{\alpha}$,

\begin{equation*}
    \ket{\psi}\otimes \ket{k_\alpha}= \left( \ket{u_{\rm H}}\ket{H} +e^{i\phi}\ket{u_{\rm V}}\ket{V}\right) \otimes \ket{k_\alpha}.
\end{equation*}
In the following we summarise how the ideal instrument matrix of Eq.~(\ref{ideal_instrument}) is realised experimentally. In a first step, a polarizing beam splitter (${\rm PBS}_1$ in Fig.~\ref{fig:exp_setup}) sends the horizontal (vertical) polarisation component of the light along path $\alpha$ ($\beta$), thereby transferring correlations between the transverse spatial and polarization degrees of freedom to the longitudinal path degree of freedom, $\ket{u_{\rm H}}\otimes\ket{H}\otimes \ket{k_\alpha} +e^{i\phi}\ket{u_{\rm V}}\otimes\ket{V}\otimes \ket{k_\beta} $. This realises the required Naimark extension. We note that the state now has the form of a Greenberger–Horne–Zeilinger (GHZ) state \cite{Greenberger2007}, with an input vector state with maximal concurrence or `vectorness'  \cite{Ndagano16,selyem19} being associated with a maximally entangled GHZ state.
The light then progresses through 
a Mach-Zehnder interferometer, which contains the half waveplates $\rm{HWP}_1$ and $\rm{HWP}_2$ in its arms and is closed by ${\rm PBS}_2$. Rotating these waveplates controls the $H : V$ polarization ratio in each arm, effectively transforming the interferometer into a partial polarizing beam splitter (PPBS) with the aid of the half-wave plate  $\rm{HWP}_3$  \cite{Ling2006,florez18}.

The state transformation realized by the waveplates ($\rm{HWP_1}$, $\rm{HWP_2}$, $\rm{HWP_3}$, $\rm{QWP}$), in the basis of the compound system $\{ \ket{H} \otimes \ket{k_\nu}, \ket{V} \otimes \ket{k_\nu}\}$ ($\nu = \alpha, \beta$), can be written as, 

\begin{eqnarray}
    \ket{H} \otimes \ket{k_\alpha} &\mapsto& (-ia\frac{\sqrt{2}}{2}(\ket{H}+ \ket{V})\otimes \ket{k_\alpha}  - i b \frac{\sqrt{2}}{2} (\ket{H}- i\ket{V}) \otimes \ket{k_\beta})e^{i\phi_\alpha},  \nonumber \\
    \ket{V} \otimes \ket{k_\alpha} &\mapsto& (ib\frac{\sqrt{2}}{2}(\ket{H}- \ket{V}) \otimes \ket{k_\alpha} - a\frac{\sqrt{2}}{2}(\ket{H} + i\ket{V}) \otimes \ket{k_\beta})e^{i\phi_\beta}, \nonumber \\
    \ket{H} \otimes \ket{k_\beta} &\mapsto& (a\frac{\sqrt{2}}{2}(\ket{H}- \ket{V}) \otimes \ket{k_\alpha} + b\frac{\sqrt{2}}{2}(-i\ket{H} + \ket{V}) \otimes \ket{k_\beta})e^{i\phi_\beta},   \nonumber \\
    \ket{V} \otimes \ket{k_\beta} &\mapsto& (-b\frac{\sqrt{2}}{2}(\ket{H}+ \ket{V}) \otimes \ket{k_\alpha} + a\frac{\sqrt{2}}{2}(\ket{H} - i\ket{V}) \otimes \ket{k_\beta})e^{i\phi_\alpha}.
    \label{statetransformn}
\end{eqnarray}
The rotation angles $\theta_i$ of $\rm{HWP}_i$ ($i$ = 1, 2) are adjusted such that $a = \sin{2\theta_1} = \cos{2\theta_2}$ and $b = \sin{2\theta_2} = \cos{2\theta_1}$, where $a$, $b$ are the coefficients in Eq.~(\ref{eqpovm}).  Effectively, the $H:V$ amplitude ratio becomes $a:b$ in path $\alpha$ and $b:a$ in path $\beta$, which can be verified by measuring the corresponding intensity ratios with a power meter.

The following polarisation optics rotate the polarisation components into the POVM basis: the fast axis of $\rm{HWP}_3$, positioned in path $\alpha$, is rotated by $67.5^\circ$ with respect to the horizontal polarization direction,  and prepares the states $\ket{\phi_1}$ and $\ket{\phi_2}$; the fast axis of the quarter-waveplate QWP, in path $\beta$, is set to $45^\circ$, preparing $\ket{\phi_3}$ and $\ket{\phi_4}$. The phases $\phi_\nu$ ($\nu = \alpha, \beta$) arise due to the optical path length of the two arms of the Mach-Zehnder interferometer. These phase differences are cancelled independently by rotating two quartz plates ${\rm QP}_1$ and ${\rm QP}_2$ around their vertical axis as part of our calibration outlined below.

The final PBS (${\rm PBS}_3$) performs a projection measurement, indicated as P in Fig.~\ref{fig:theoryfig}, separating $\ket{\phi_1}$ and $\ket{\phi_2}$ in path $\alpha$ and $\ket{\phi_3}$ and $\ket{\phi_4}$ in path $\beta$.
The intensity profiles of the corresponding POVM elements are recorded in different quadrants of a CMOS camera, as shown in the subset in Fig.~\ref{fig:exp_setup}. 
We confirm that the unitary transformations expressed in Eqs.~(\ref{statetransformn}) are identical to the desired unitary transformation matrix $U$ shown in Fig.~\ref{fig:theoryfig} and given in the Supplementary Documentation Eq.~(\ref{unitarymatrix1}).

One of the most crucial steps in the calibration of our setup is the elimination of unwanted phase shifts with ${\rm QP}_1$ and ${\rm QP}_2$. A suitable technique was discussed in \cite{Ling2006} for homogeneously polarized light. In contrast to their application, here we need to cancel phase shifts across extended beam profiles to achieve interferometric stability, which increases the experimental challenge significantly.  \new{The inherent phase sensitivity of the Mach-Zehnder interferometer meant that our system required frequent realignment. Calibration measurements were performed before and after data runs to ensure stability.} 

\begin{figure}[H]
    \centering
    \includegraphics[scale=1.0]{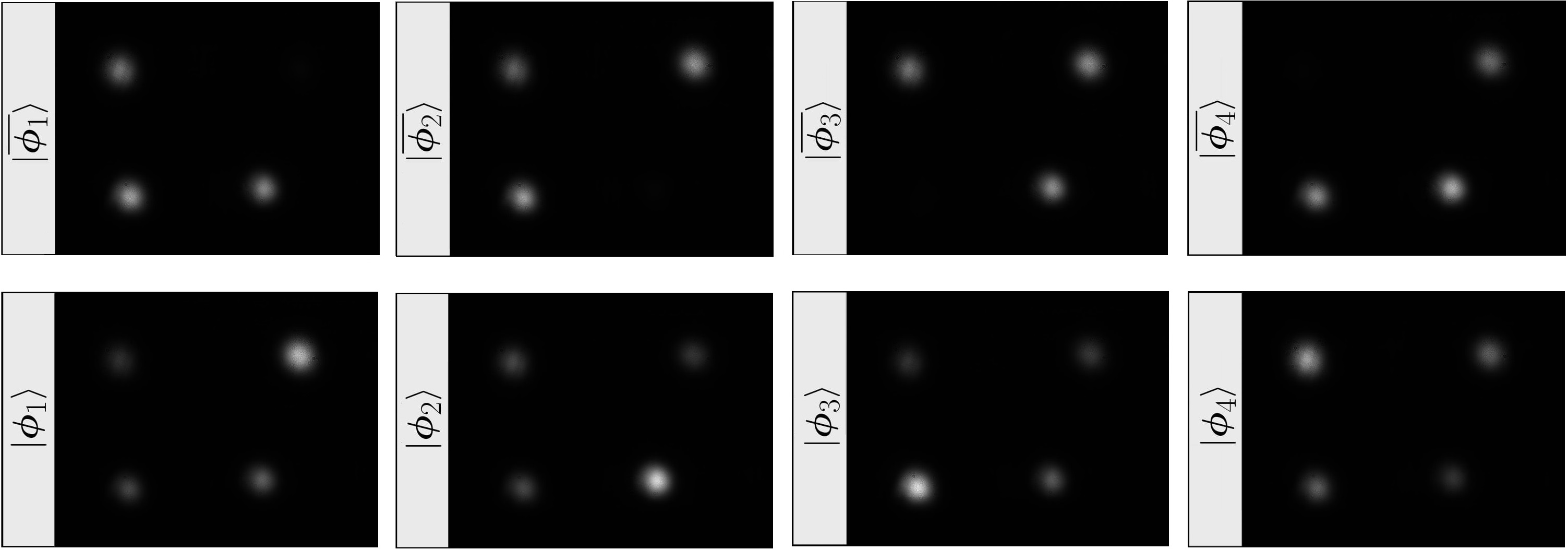}
    \caption{Calibration measurements: Experimental intensity measurements $I_i$ ($i \in \{1,2,3,4\}$) as recorded on the CMOS camera in the POVM setup for homogeneously polarised input light. Measurements for light prepared  in the orthogonal modes $\ket{\bar{\phi}_i}$ and for light prepared in the POVM states $\ket{\phi_i}$.}
    \label{fig:plusminusstate}
\end{figure}

Using suitable multiplexed holograms on our DMD, we prepare homogeneously polarized light in the POVM states $\ket{\phi_i}$ from Eq.~(\ref{eqpovm}) as well as in states $\ket{\overline{\phi}_i}$ orthogonal to our POVM states,

\begin{eqnarray}\label{eqpovmnot}
    \ket{\overline{\phi}_1}= b\ket{H} - a\ket{V}, \qquad  
    & \ket{\overline{\phi}_2}= b\ket{H} + a\ket{V}, \nonumber \\
    \ket{\overline{\phi}_3}= a\ket{H} - ib\ket{V}, \qquad
    & \ket{\overline{\phi}_4}= a\ket{H} + i b\ket{V}. 
    \label{statetetrahedron}
\end{eqnarray}
For a perfectly calibrated system, the measured intensities for each state $\ket{\overline{\phi}_{i}}$ ($i \in \{1,2,3,4\}$) should be $I_i=0$, with equal intensities in all other quadrants. 
Conversely, light prepared in the POVM states $\ket{\phi_{i}}$ should yield a maximum intensity ${\rm I}_i$, as the state is projected onto itself, with lower equal intensities for all other POVM elements. 
We adjust iteratively ${\rm QP}_1$ and ${\rm QP}_2$ to approximate these outcomes. Experimental measurements after calibration are shown in Fig.~\ref{fig:plusminusstate}, agreeing qualitatively with the desired outcomes, but showing small quantitative differences that indicate a remaining imbalance in the phase compensation.  

To assess the operation of our POVM tomography quantitatively we calculate the experimental instrumentation matrix of our system and compare it to the ideal instrument matrix Eq.~(\ref{ideal_instrument}).
We recall that the instrument matrix links the normalised Stokes vector ${\bf S}^{\rm (N)}$ of the input light to the normalised intensities ${\bf I}^{\rm (N)}$ at the output of the POVM measurement, as expressed by Eq.~(\ref{instrument_eqn}).  Inverting this equation allows us to identify the experimental instrument matrix.  
The simplest way to do this is to use the 4 POVM states as the generated input beams passing through the calibrated POVM measurement system, and recording the resulting intensities ${\bf I}^{\rm (N)}$, as shown in the bottom row of Fig.~\ref{fig:plusminusstate}. Any changes to the state of the input beam induced by the experimental setup are reflected as a small deviation in the value of the entries of the experimental instrumentation matrix $\Pi_{\rm{exp}}$ when compared to the ideal matrix listed in (\ref{ideal_instrument}). A typical matrix is given by,

\begin{equation}
    \label{exp_instrument}
    \Pi_{\rm{exp}} = \frac{1}{4}\begin{pmatrix}
        1.05 & 0.77\sqrt\frac{1}{3} & 1.10\sqrt\frac{2}{3} & 0.05 \\
        1.02 & 1.21\sqrt\frac{1}{3} & -0.88\sqrt\frac{2}{3} & -0.04 \\
        1.06 & -1.24\sqrt\frac{1}{3} & -0.01 & -0.93\sqrt\frac{2}{3} \\
        0.88 & -0.74\sqrt\frac{1}{3} & -0.03 & 0.89\sqrt\frac{2}{3}
    \end{pmatrix},
\end{equation}
where we have used a notation that highlights the deviation from the theoretical instrument matrix $\Pi$ from Eq.~(\ref{ideal_instrument}). It is worth noting that tomography can be performed as long as the instrument matrix permits an invertible mapping between the POVM elements and the state as parametrized e.g.~via the Stokes vector. The discrepancy between theoretical and measured instrument matrix simply means that the experimental setup projects onto slightly different states than the intended MIC-POVM elements $\ket{\phi_{i}}$.

We assume that the deviation from the desired instrument matrix arises mainly due to imperfect phase compensation and potentially remaining imbalanced optical activity experienced along the different beam paths. Another possible source of error could arise from inhomogeneous effects across the beam profile. This latter effect could be compensated by evaluating a spatially resolved experimental instrument matrix, \new{however we find that this} did not improve the fidelity. 
When analysing our POVM setup we therefore evaluate the performance using the experimental instrument matrix Eq.~(\ref{exp_instrument}).


\section{Analysing performance and discussion}

\begin{figure}[H]
    \centering
        \includegraphics[width=0.95\linewidth]{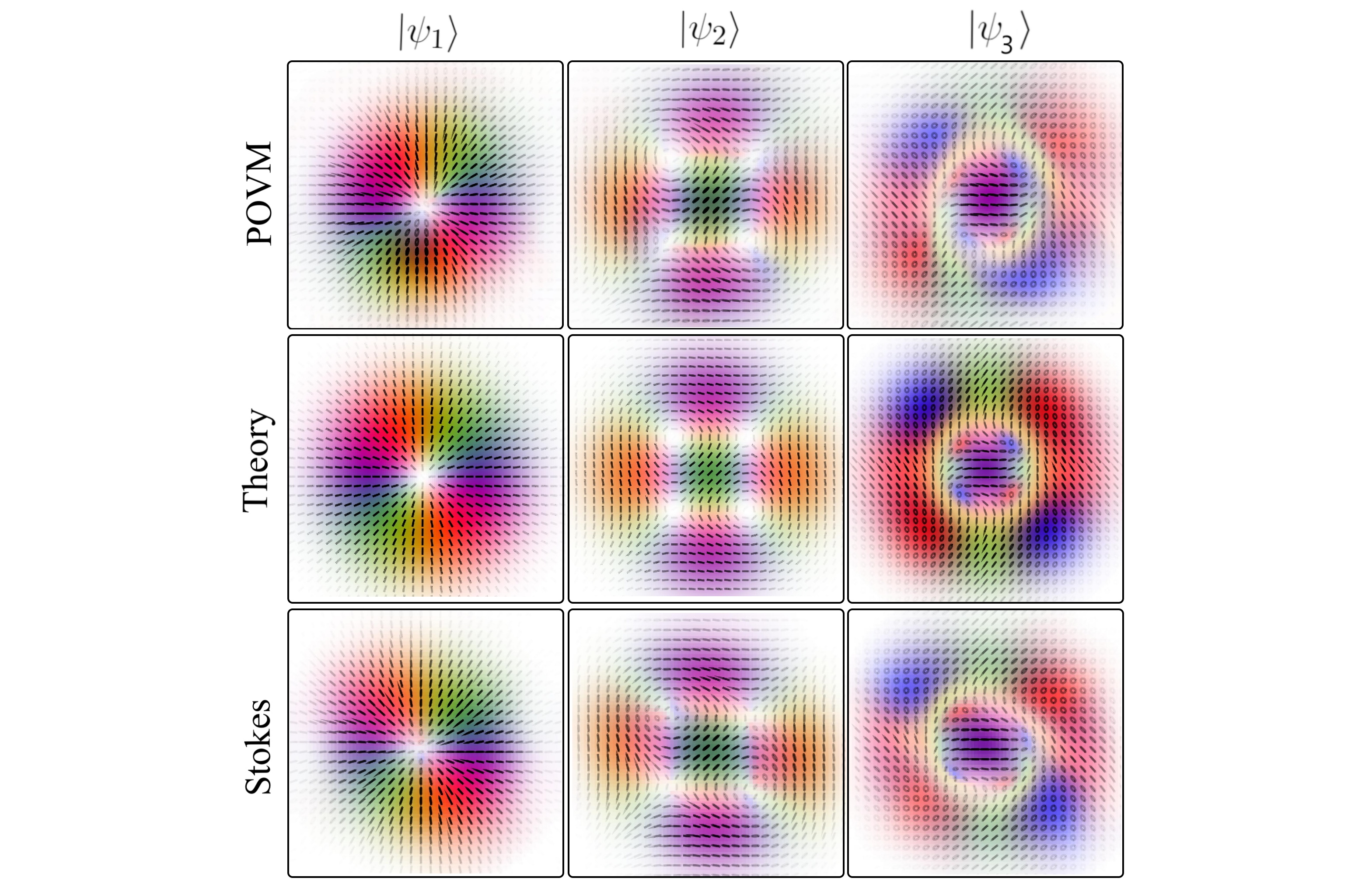}
    \caption{Qualitative verification of POVM tomography: Polarization distributions of the three test beams $\ket{\psi_{1}}, \ket{\psi_{2}}$ and $\ket{\psi_{3}}$, obtained by POVM tomography (top), simulation (middle) and Stokes tomography (bottom). We use the color scheme indicated in Fig.~\ref{fig:poincare} to characterize the polarization state, and beam opacity for the intensity profile.} \label{fig:polarization}
\end{figure}

We have assessed the performance of our POVM tomography technique for various vector beams of the general form given in Eq.~(\ref{state}), three of which we present in this paper,
\begin{eqnarray}\label{teststates}
    \ket{\psi_1} & = & \ket{\rm{HG}_{1,0}}\ket{\rm{H}}+\ket{\rm{HG}_{0,1}}\ket{\rm{V}}, \nonumber \\   
    \ket{\psi_2} & = &  \ket{\rm{HG}_{0,2}}\ket{\rm{H}}+\ket{\rm{HG}_{2,0}}\ket{\rm{V}}, \\
    \ket{\psi_3} & = & \ket{\rm{LG}_1^0}\ket{\rm{H}}+\ket{\rm{LG}_0^2}\ket{\rm{V}}.\nonumber
\end{eqnarray}
Here $\ket{{\rm HG}_{n,m}}$ are Hermite-Gaussian modes with horizontal and vertical mode indices $n$ and $m$, and $\ket{{\rm LG}_p^\ell}$ describe Laguerre-Gaussian modes with radial mode number $p$ and azimuthal mode number $\ell$. The theoretical intensity and polarization profiles of these beams are shown in the middle row of Fig.~\ref{fig:polarization}. The first beam, $\ket{\psi_1}$, has a radial polarization profile, the second beam, $\ket{\psi_2}$, is a non-radially symmetric beam with four-fold symmetry, and the final beam, $\ket{\psi_3}$, is a Poincar\'e beam, that carries a net OAM of $1\hbar$ per photon, with different LG beams in the horizontal and vertical polarization components. 

When assessing the performance of our tomography, we need to differentiate between errors that occur during the experimental state generation and those due to the detection process. While homogeneous polarization states can be generated with near perfect accuracy, the preparation stage will lead to a deviation of the actual test beam from the desired polarization pattern. In order to evaluate the measurement process itself, we compare POVM tomography to conventional six-measurement Stokes tomography.

We verify POVM tomography by comparing the obtained polarization maps against the theoretical prediction for the structured beams Eq.~(\ref{teststates}) and those obtained by Stokes polarimetry, and quantify this by giving the respective angular accuracy in terms of the average angular deviation between the Stokes vectors on the Poincar\'e sphere.

\begin{figure}[H]
    \centering
        \includegraphics[width=0.85\linewidth]{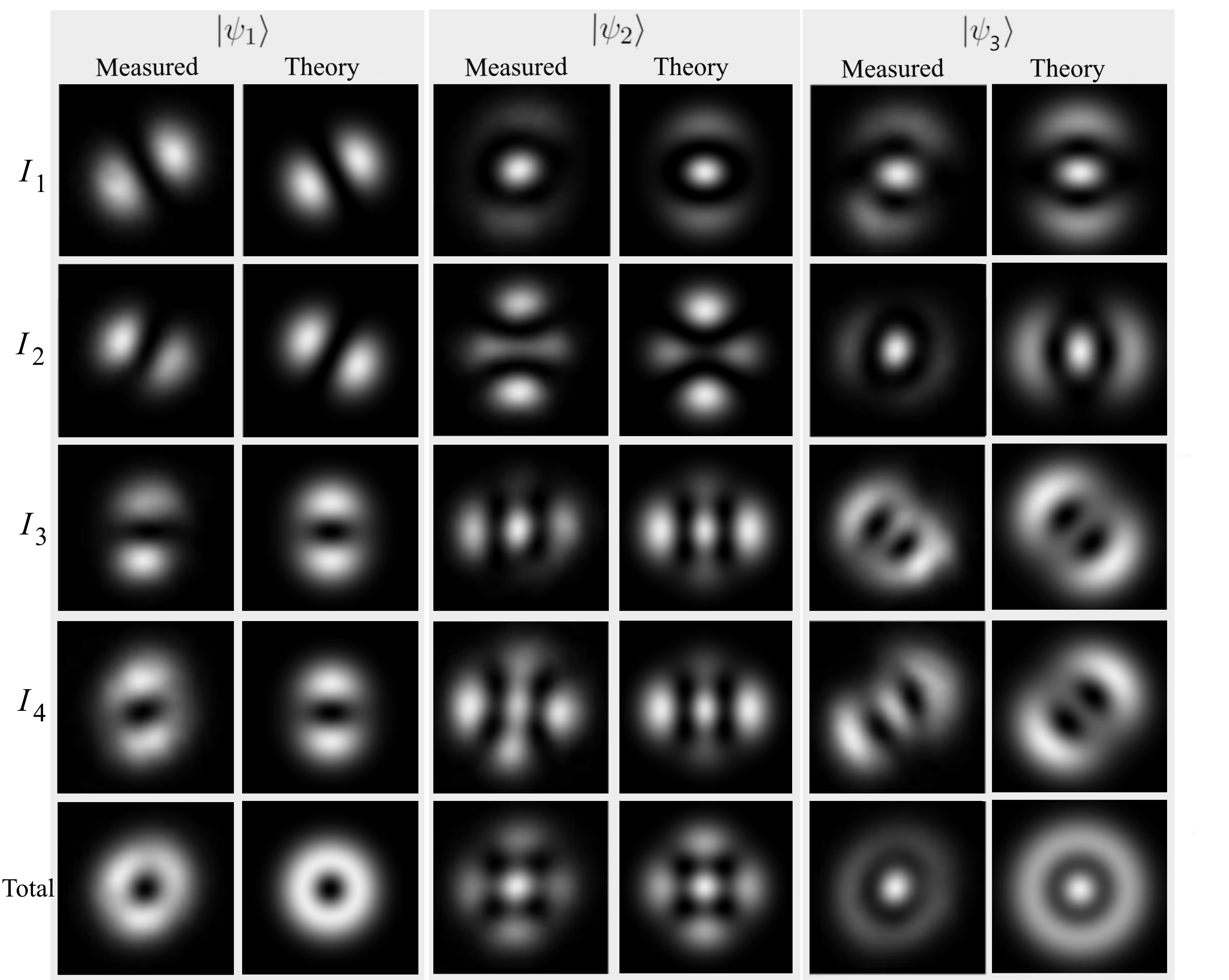} \caption{Total intensity and measurements of the four POVM elements for $\ket{\psi_1},\ket{\psi_2},\ket{\psi_3}$.
        } 
    \label{fig:intensities}
\end{figure}

The data obtained from our POVM single-shot measurement are the intensity profiles shown in the first column of Fig.~\ref{fig:intensities}a) to c) for our three test beams, and recorded in the four quadrants of a CMOS camera. \new{In these images, the background is subtracted from the camera images, obtained as the average light level over a small camera section far from the beam areas.}  
We have also removed noise by applying a low-pass Fourier filter and identified their respective centers with moment analysis. Overlapping the intensities of the four POVM components results in the total intensity distribution (bottom row), which serves as a control that the individual intensities profiles are properly aligned. 
We compare these intensity measurements to the intensity patterns predicted for the theoretical beams, shown in the second columns respectively of Fig.~\ref{fig:intensities}a) to c), and generally find good qualitative agreement. 

From the measured intensity data we can fully reconstruct the vector beam, and in particular the spatially varying Stokes vectors by inverting Eq.~(\ref{instrument_eqn}) for the experimental instrument matrix Eq.~(\ref{exp_instrument}). The resulting polarization maps are shown in the first row of Fig.~\ref{fig:polarization}, and compared with the theoretical prediction (second row) and the maps reconstructed from spatially resolved six-measurement Stokes tomography (bottom row).

We find that the overall topology is accurately reproduced, but suffers from some degree of distortion
, indicating a reasonable (but not perfect) preparation fidelity. We discuss the origin of the observed distortions in the later part of this section.

\begin{figure}[H]
    \centering
        \includegraphics[width=1\linewidth]{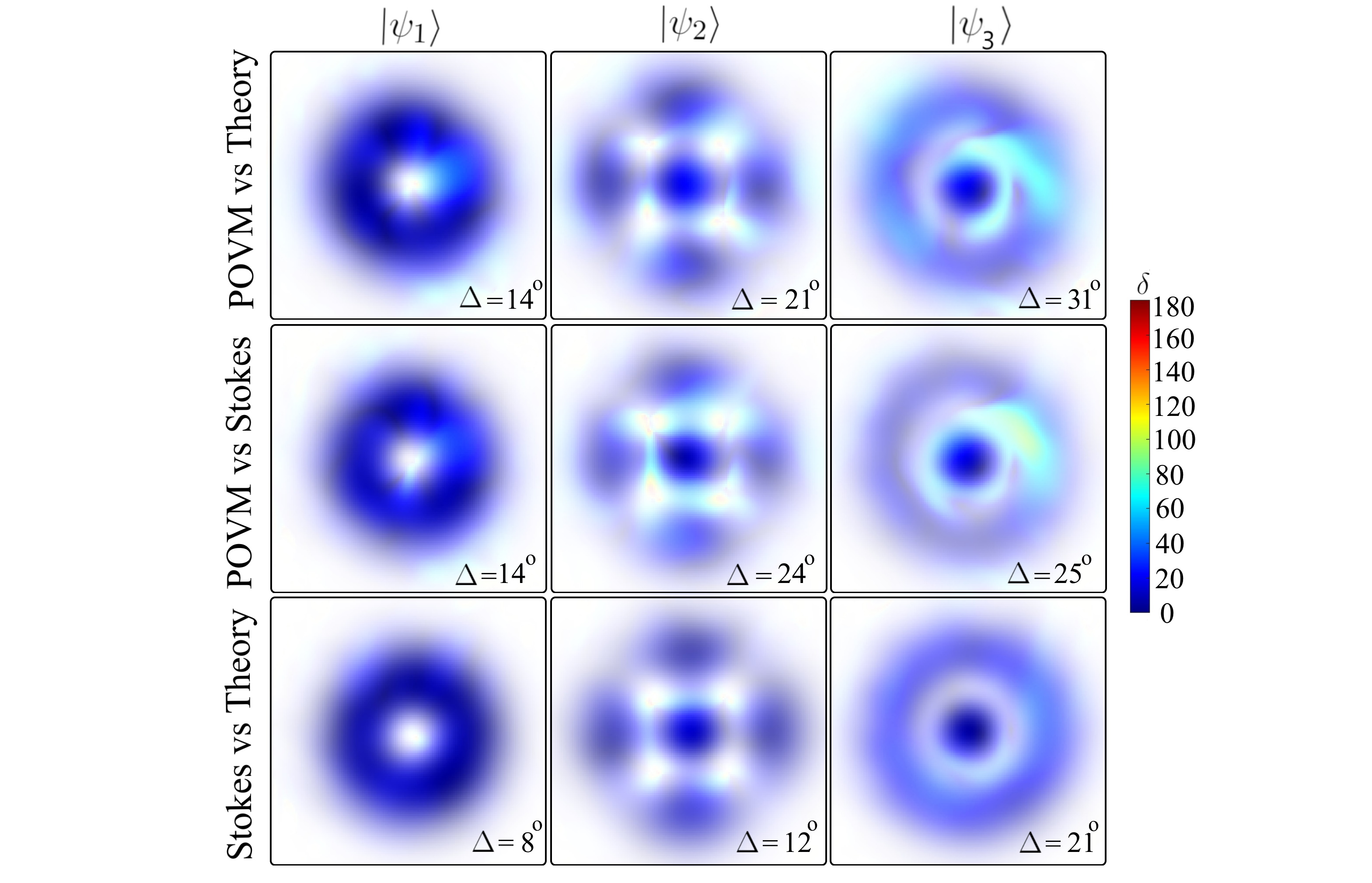} 
    \caption{Quantitative evaluation of POVM tomography: For three test beams we show the angular deviation distribution $\delta$, and its intensity weighted average $\Delta$, comparing the polarization profiles between POVM tomography and theory (top row), POVM and Stokes tomography (middle) and between Stokes and theory (bottom row). Blue indicates agreement, and red discrepancies.}
    \label{fig:heatmaps}
\end{figure}

In order to provide a quantitative assessment of our reconstructed beams, we calculate the spatially resolved angular deviation angle $\delta$ \cite{Ryan2019} of the polarization state on the Poincar\'e sphere, pairwise comparing POVM tomography, Stokes tomography and theoretical prediction.  The deviation angle,
\begin{equation}
    \delta= \rm{cos}^{-1}\left( \frac{\textbf{S}_{123}\cdot\textbf{S}_{123}^{'}}{\mid\textbf{S}_{123}\mid\mid\textbf{S}^{'}_{123}\mid}\right)  
\end{equation}
can be interpreted as the angle between the polarisation states on the Poincar\'e sphere as defined by the reduced Stokes vectors $\textbf{S}_{123}$ and $\textbf{S}^{'}_{123}$ obtained from different methods.
An angular deviation of 0 would indicate that the reconstructed polarization states match perfectly, whereas orthogonal states are separated by $180^{\circ}$, and $\delta = \varphi$ indicates e.g. a linear polarization rotation of $\varphi/2$. Fig.~\ref{fig:heatmaps} shows the angular deviation distributions for our test beams, as well as the light intensity-weighted angular deviation averaged over the entire beam, denoted by $\Delta$. For both POVM and Stokes tomography, regions of larger angular deviation often correspond to regions of lower light intensity, where noise plays a larger role. 

Quantitative analysis confirms that POVM tomography allows the reconstruction of polarization profiles from the minimum number of required measurements - however with slightly less accuracy than Stokes tomography. 
This difference is not surprising, given the increased experimental complexity of the POVM setup on one hand, and the overcomplete measurements of the Stokes tomography, which provide redundant information which generally reduces errors.

The discrepancies between Stokes tomography and theoretical profiles can arise from errors in the preparation process and from the measurement process, e.g. due to misalignments between the polarization components.   
Naturally, any imperfections in the preparation process will also impact the POVM tomography. Regarding the POVM measurement process, we noted before that the main experimental challenge was to compensate the phase difference acquired in the Mach-Zehnder interferometer via the quartz plates ${\rm QP}_{1,2}$ across the whole beam profile. 
\new{More generally, our POVM setup is more complicated, including longer beam paths which incorporate different optical elements, which may affect the polarization distribution of the beam prior to the projection.} 
Finally, we note that our experimental alignment procedure favors Stokes tomography, as we tune the multiplexed hologram displayed on the DMD with homogeneous test polarizations identified via the Stokes measurement system. This allows us to adjust for polarization transformations along the beam path to the Stokes detection system, but currently not to the POVM polarimeter.


\section{Conclusion}

Our results demonstrate in a proof-of-principle experiment that full characterization of arbitrary vector vortex beams in a single-shot measurement is possible. The achieved accuracy is mainly limited by distortions of the polarization patterns upon propagation. This is observed even for the experimentally simpler process of Stokes tomography, and becomes more noticeable for beams with more complicated beam structures including Poincar\'e beams. \new{It is interesting to note that transverse variations in phase, e.g.~due to differential aberrations, acquired Gouy phases or effects of optical elements, leave the overall topology intact}, effectively redistributing the polarization structure across the beam profile.  It would be interesting to test this in a full investigation of retrieving the vector quality factor \cite{Nape2021} and optical skyrmion number \cite{Gao2020,Shi2021}, to affirm the invariance of vectorial structured light due to dephasing and aberrations. 

In future work we aim to reduce transverse \new{variations in phase} by replacing the Mach-Zehnder with a Sagnac configuration. Further developments include replacing the current spatially resolved detection with projections onto pre-defined spatial modes, allowing detection via single-photon detectors rather than camera, and thus boosting efficiency for applications in the quantum regime.  

In the classical regime, our system could serve as a spatially resolved one-shot analysis for photo-active materials. Conventional methods to identify the birefringent or dichroic properties of a substance rely on identifying its Mueller matrix. This typically requires monitoring the response of the medium to the six input polarizations $\rm{I}_i({\bf r}_{\bot})$ for $i \in {\{ \rm H,V,D,A,R,L\}}$ sequentially, requiring a total of 36 measurements.  By employing POVM tomography, this could be reduced to 16 measurements realised in 4 sequential measurements. For homogeneous optical activity, e.g.~of a liquid or gas, this could be further reduced by using the spatial degree of freedom to replace the sequential measuremnts. Exposing the substance to a Poincar\'e beam would allow testing the response across all polarizations simultaneously, generating a single-shot method for testing optical activity.  This could be of importance for photo-sensitive materials that cannot be exposed to prolonged test light, of for scenarios that require monitoring dynamic changes of optical activity.


\section*{Data Availability}
The datasets generated and analyzed during the study are available from Enlighten repository at \cite{dataset}.


\section*{Conflict of Interest Statement}

The authors declare that the research was conducted in the absence of any commercial or financial relationships that could be construed as a potential conflict of interest.


\section*{Author Contributions}

M.A.A and C.M.C have taken and analysed the data, H.J has been involved in early data gathering and analysis, S.P has built the initial setup and together with S.C overseen the theoretical analysis, S.P and S.F.-A have conceived and led the work. All authors have contributed to writing the manuscript.


\section*{Funding and Acknowledgments}

M.A.A is supported through a QuantIC scholarship from EPSRC (EP/M01326X/1). C.M.C acknowledges financial support from the Royal Society through a Newton International fellowship (NIFR1192384). S.P acknowledges the Coordena\c{c}\~ao de Aperfei\c{c}oamento de Pessoal de N\'ivel Superior (CAPES-PRINT) for supporting his stay as visiting professor at the University of Glasgow during the preparation of this work. S.C is supported by a Leverhulme Fellowship (RF-2020-397).
S.F.-A gratefully remembers initial discussions with Najmeh Tabebordbar, Isaac Nape and Andrew Forbes at the very beginning of this project. \\

\bibliography{manuscript}


\section*{Supplemental Material}

\setcounter{equation}{0}
\renewcommand{\theequation}{S\arabic{equation}}

Here we give more detail on how the Naimark extension of the POVM is realised, in order to implement the four outcome POVM measurement described in the text. The unitary transformation applied to the input state, written in the extended basis reads,

\begin{equation}
      U = \sum_{\substack{i,j=1,2;\\ \mu,\nu=\alpha,\beta}} U_{\mu ij \nu} \ket{k_\mu}\ket{e_i} \bra{e_j}\bra{k_\nu},
        \label{unitary1}
\end{equation}
where $\bra{e_1}$ and $\bra{e_2}$ denote orthogonal polarisation states and $\bra{k_\alpha}$ and $\bra{k_\beta}$ different paths. Here $U_{\mu ij \nu} = \bra{k_\mu}\bra{e_i}U\ket{e_j}\ket{k_\nu} $ are the elements of the matrix $U$,

\begin{equation}
    U=\frac{\sqrt{2}}{2}
    \begin{pmatrix}
        -ia &  ib &  a & -b\\  
        -ia & -ib & -a & -b \\ 
        -ib & -a & -ib & a \\     
        -b & -ia & b& -ia
    \end{pmatrix},
    \label{unitarymatrix1}
\end{equation}
written in the basis of the extended Hilbert space $\{\ket{e_1}\ket{k_\alpha}, \ket{e_2}\ket{k_\alpha}, \ket{e_1}\ket{k_\beta}, \ket{e_2}\ket{k_\beta} \}$ in this order.
The final projective measurement $P$ can be realised using a polarizing beamsplitter combined with a CMOS camera for spatially resolved detection.

For an input state defined in the extended state space,
\begin{equation}
\label{input}
\ket{\psi}\otimes \ket{k_\alpha}= \left( \ket{u_{H}}\ket{H} +e^{i\phi}\ket{u_{V}}\ket{V}\right) \otimes \ket{k_\alpha},
\end{equation}
the probability that a photon is detected in one of the exits of the interferometer, 
identified by the path $\nu$ and polarisation $i$, 
is therefore given by \cite{barnett09},
\begin{equation}
    \bra{\psi}\hat{\pi}_{i\nu}\ket{\psi} = \lvert  \bra{k_\nu}\bra{e_i}U\ket{\psi} \ket{k_\alpha}  \rvert^{2},
    \label{probability}
\end{equation}
where $\hat{\pi}_{i\nu} = \bra{k_\alpha}U^{\dagger}\ket{e_i}\ket{k_\nu}\bra{k_\nu}\bra{e_i}U\ket{k_\alpha}$ are the POVM elements, which using Eq.~(\ref{unitary1}) may be expressed as,

\begin{equation}
    \hat{\pi}_{i\nu} =  \sum_{j^{\prime},j=1,2} U_{\nu ij^{\prime}\alpha}^{\ast} U_{\nu i j\alpha} \ket{\hat{e}_{j^{\prime}}} \bra{\hat{e}_j}.
    \label{povm1}
\end{equation}

It is then readily verified that choosing $U$ as given in Eq.~(\ref{unitarymatrix1}) 

and using the notation transformation $\hat{\pi}_1=\hat{\pi}_{1\alpha}$, $\hat{\pi}_2=\hat{\pi}_{2\alpha}$, $\hat{\pi}_3=\hat{\pi}_{1\beta}$, $\hat{\pi}_4=\hat{\pi}_{2\beta}$, we obtain $\hat{\pi}_i=\frac{1}{2}\ket{\phi_i}\bra{\phi_i}$ for $i = 1,\dots,4$, with the projective measurements given by their respective expectation values, 
\begin{equation} P_{i}=\bra{\psi} \hat{\pi}_i \ket{\psi}.
\end{equation}

\end{document}